\documentclass[12pt]{article}

\def\be{\begin{equation}}
\def\ee{\end{equation}}
\def\bea{\begin{eqnarray}}
\def\eea{\end{eqnarray}}

\def\3vd{\rangle{\hspace{-0.18em}\longrightarrow}{\hspace{-0.65em}^\mid}}
\def\4vd{\rangle{\hspace{-0.15em}\longrightarrow}}

\def\sw{\!\wedge\!}

\def\2d{\dot{2}}
\def\1d{\dot{1}}
\def\3d{\dot{3}}
\def\e{\epsilon}
\def\jsl{J\hspace{-0.78em}\diagup}
\def\esl{\e\hspace{-0.4em}/}
\def\sl{\hspace{-0.6em}/}

\def\kk{\kappa}

\def\cdt{\cdots}

\def\m{\mu}
\def\n{\nu}
\def\mm{\mid}
\def\1q{Q+1}
\def\2q{Q+\kk(1,2)}
\def\3q{Q+\kk(1,3)}

\usepackage{graphicx}
\usepackage{bm}
\usepackage{amsmath}
\usepackage{amssymb}

\begin{document}

\begin{flushright}
hep-th/0609083
\end{flushright}

\pagestyle{plain}

\begin{center}
\vspace{2.5cm}

{\Large {\bf Amplitudes With Different Helicity Configurations Of Noncommutative QED}}

\vspace{1cm}

Abolfazl Jafari

\vspace{.5cm}

{\it Institute for Advanced Studies in Basic Sciences (IASBS),\\
P. O. Box 45195, Zanjan 1159, Iran}

\vspace{.3cm}

\texttt{jabolfazl@iasbs.ac.ir}

\vskip .5 cm
\end{center}

\begin{abstract}
The amplitudes of purely photonic and photon--2-fermion processes of
noncommutative QED (NCQED) are derived for different helicity configurations of photons.
The basic ingredient is the NCQED counterpart of Yang-Mills recursion relations by means of
Berends and Giele. The explicit solutions of recursion relations for NCQED photonic processes
with special helicity configurations are presented.
\end{abstract}

\vspace{2cm}

\noindent {\footnotesize Keywords: Noncommutative Geometry, Noncommutative Field Theory}\\
{\footnotesize PACS No.: 02.40.Gh, 11.10.Nx, 12.20.-m}
\date{\today}

\newpage

\section{Introduction}
The calculation of cross sections for the production of many particles
in high-energy collisions is restricted by the technical
difficulties associated with the evaluation of the corresponding
multi-particle Feynman diagrams. In contrast to the usual
approach, one can express the amplitudes in terms of positive and
negative chirality spinors ($\lambda$ and $\tilde{\lambda}$) and
their corresponding products. It is known that the amplitudes associated
to special helicity configurations have very simple expressions
in terms of these spinor products \cite{stefan, giele, cachazo, ozeren, str}.

Much significant progresses have been made in the last years with the
development of new techniques, principally the Maximally Helicity Violating (MHV) rules
introduced by Cachazo, Svrcek and Witten (CSW) \cite{csw}, and
also the the recursion relation among amplitudes by Britto-Cachazo-Feng (BCF) \cite{bcf}.
CSW introduced a novel diagrammatic technique, MHV rules, in which maximally
helicity violating amplitudes are used as vertices in a scalar perturbation theory, for which
\emph{the vertices are connected by scalar propagators $\displaystyle{\frac{1}{p^{2}}}$}.
This arrangement vastly reduces the number of diagrams that must
be evaluated relative to the traditional Feynman rules case.
Although the original CSW paper dealt only with purely gluonic
amplitudes, the formalism has been successfully extended to
include quarks Higgs and massive gauge bosons.

The focus of attention has been on developing and exploiting
techniques for QCD scattering processes, since such amplitudes are
needed to estimate multi-jet cross sections at hadron colliders
\cite{ozeren,kleiss, wu, zhanf, kunst}. Indeed, the MHV rules were
specifically developed for and applicable to massless Yang-Mills
field theory. As mentioned, particular helicity amplitudes in Yang-Mills theory
take on unexpectedly simple forms \cite{witten, kosower, dixon, anton}.
At tree level for example, purely gluonic colure-ordered
scattering amplitudes can be summarized as follows
\bea
\varepsilon(1\pm,2+,\cdots,n+) &=& 0\nonumber\\
\varepsilon(1-,2-,\cdots,n+)&=&\frac{\langle 12\rangle^{4}}
{\Pi_{k=1}^{n}\langle  k\   k+1 \rangle }
\eea
So amplitudes with all the gluons having the same helicity vanish, as do those with
only one gluon having a different helicity to the others. The
second case above therefore corresponds to the maximally helicity
violating amplitudes. Their simple forms were first
conjectured by Parke and Taylor \cite{park}, and later proven by
Berends and Giele using a recursive technique \cite{giele}. The amplitude
$$
\varepsilon(1+,2+,\cdots,i-,\cdots,j-,\cdots,n+)
$$
is called a ``mostly plus" MHV amplitude, for obvious reasons. Its
'mostly minus' counterpart, which has two positive helicities and
the remainder negative, is called an $\overline{\rm MHV}$
amplitude, and can be obtained by interchanging $\langle  ij
\rangle \rightarrow[ij]$. Another example is
\bea
\varepsilon(\bar{f}+,f-,1+,2+,\cdots,n+)& =& 0\nonumber\\
\varepsilon(\bar{f}+,f-,1+,2+,\cdots,i-,\cdots,n+)& =&
\frac{2^{\frac{n}{2}}e^{n}\langle  f\bar{f} \rangle ^{n-2}\langle
fi \rangle ^{3} \langle  \bar{f}i \rangle }{\Pi_{k=1}^{n}\langle
fk \rangle \langle  \bar{f}k \rangle }
\eea
Here $i^{+}$ denotes a positive helicity photon with momentum $p_{i}$ and $f,\bar{f}$
denote fermion and anti-fermion respectively. This is the
fundamental MHV amplitude in QED, and as before it consists of
only a single term. The factor $e^{n}$ is the gauge coupling
constant \cite{ozeren}.

On the other hand, recently a great interest has been appeared to
study field theories on spaces whose coordinates do not commute.
These spaces, as well as the field theories defined on them, are
known under the names of noncommutative spaces and theories. In
contrast to U(1) gauge theory on ordinary space-time, as we
briefly review in next section, noncommutative version of theory
is involved by direct interactions between photons. Interestingly
one finds the situation very reminiscent to that of non-Abelian
gauge theories, and then the question is whether the techniques
developed for non-Abelian theory purposes can be used for
noncommutative QED case too. In particular, the same question may
arise for the recursive relation techniques
\cite{9908142,reviewnc, phen-nc1, phen-nc2, phen-nc3, mahajan, morencqed, jab}.

In \cite{jafari2} we present NC-photonic recursion relation by means
of Berends and Giele. There we show that, though in NCQED, in contrast
to momentum independent color factors of ordinary Yang-Mills theory,
the vertex functions depend on momenta, one can derive the recursive
relations for currents. In particular, the cosine of a very special
combination of momenta comes with the QCD-like currents in recursive relations.

In this paper, we investigate whether similar technique exists for
noncommutative QED (NCQED). We use Weyl-van der Waerden spinors
\cite{stefan, giele, kleiss,kosower, dixon} and its related calculus
with spinorial formalism. The amplitudes of purely photonic and photon--2-fermion processes of
noncommutative QED (NCQED) are derived for different helicity configurations of photons.
The basic ingredient is the NCQED counterpart of Yang-Mills recursion relations by means of
Berends and Giele. The explicit solutions of recursion relations for NCQED photonic processes
with special helicity configurations are presented.

\section{MHV For Photonic Processes Of NCQED}
The recursion relations for noncommutative photons are presented in
\cite{jafari2}. The NC-photonic recursion relation consist two parts.
One, as counterpart of color factors of ordinary Yang-Mills theory, consists
the cosine of a very special combination of photon's momenta and
noncommutativity parameter. The other part is the same of the color-independent part
of ordinary Yang-Mills theory, that consists just the momenta.
It is shown that since there is no inner product between the cosine part
(and the momenta it consists), and as the currents involved in ordinary Yang-MIlls and that of NCQED are
similar, we can use all properties of gluonic currents given by \cite{giele}
in case of NCQED too. The basic tool we use is the so-called spinor formalism.
We give expressions for purely photonic and photon--2-fermion
processes of NCQED.

\begin{figure}
\begin{center}
\includegraphics[width=0.8\columnwidth]{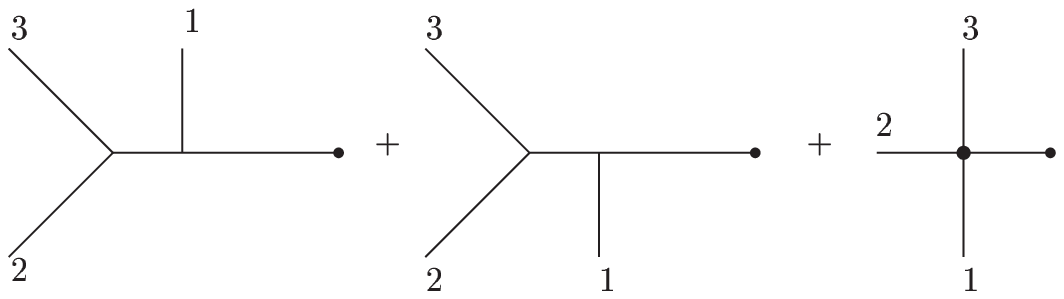}
\caption{3-photon case.} \label{photons}
\end{center}
\end{figure}

\begin{figure}
\begin{center}
\includegraphics[width=0.9\columnwidth]{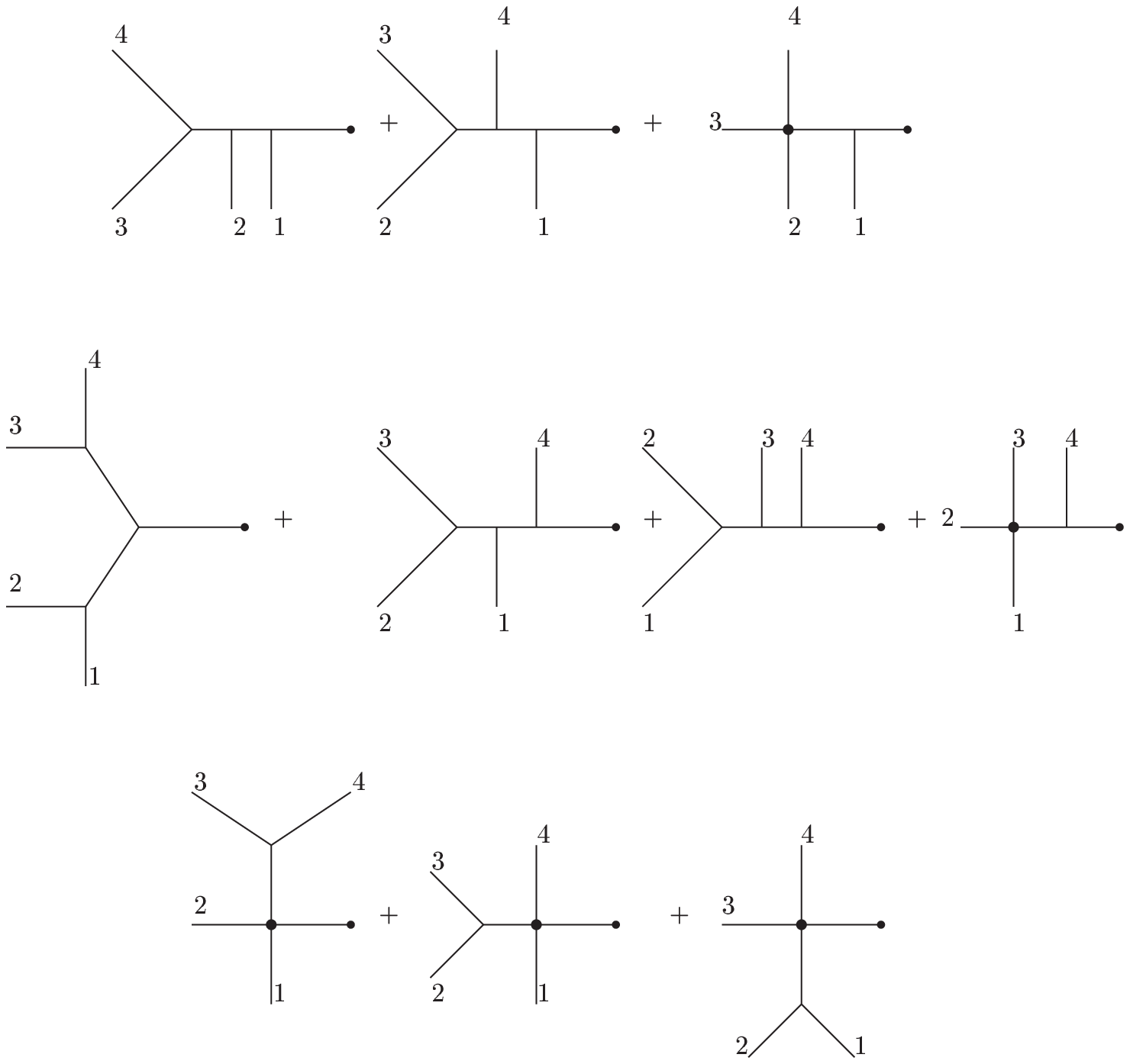}
\caption{4-photon case.} \label{feynman}
\end{center}
\end{figure}

First let us introduce the notations; see \cite{jafari2}. We use
\bea
&& \frac{1}{2}k^{a}_{i}\theta_{ab}k^{b}_{j}\rightarrow i\sw
j\nonumber\\&& i\sw klm\cdots=i\sw( k+l+m+\cdots)
\eea
\noindent with $\theta_{ab}$ as the noncommutativity parameter. Also we use the symbol ``~;~" as
\bea
ijkl\cdots ;=i\sw (j+k+l+\cdots)+j\sw(k+l+\cdots)+k\sw (l+\cdots)+\cdots
\eea
\noindent and also
\bea
\kk(1,m)\kk(m+1,k)\kk(k+1,n)\cdt
;&=&\kk(1,m)\sw(\kk(m+1,k)+\kk(k+1,n)+\cdt)\nonumber\\
&+&\kk(m+1,k)\sw (\kk(k+1\cdt n)+\cdt)+\cdt
\eea
\noindent with $\kk^{\mu}(i,j)=\sum_{l=i}^{j}l^{\mu}$. The NCQED recursive
relation for purely photonic processes is obtained to be
\cite{jafari2}
\bea
\widehat{J}_{\xi}(1,2,\cdots,n)=(-1)^{n}(ie)^{n-1}\sum_{P(1,\cdots,n)}
\;C(123\cdots n;+\frac{n-3}{2}\pi)J_{\xi}(1,\cdots,n) \label{22}
\eea
\noindent in which $e$ as coupling constant, and
\bea
C(123\cdots n;+\frac{n-3}{2}\pi)&=&
\cos \Big(1\sw (2+\cdots +n)+ 2\sw (3+\cdots +n)+\cdots \nonumber\\
&+& (n-2)\sw((n-1)+n)+ (n-1)\sw n+\frac{n-3}{2}\pi\Big) \eea The
current $J_\xi$ is that of \cite{giele, jafari2}
\bea
J_\xi(1,\cdots,n)&=&\frac{1}{\kappa^2(1,n)}
\Big{(}\sum_{m=1}^{n-1}[J(1,\cdots,m),J(m+1,\cdots,n)]_\xi\nonumber\\&+&
\sum_{m=1}^{n-2}\sum_{k=m+1}^{n-1}\{J(1,\cdots,m),J(m+1,\cdots,k),J(k+1
,\cdots,n)\}_\xi\Big{)}
\eea
\noindent in which
\bea
[J(1,\cdt,m),J(m+1,\cdt,n)]_{\xi}\!\!&=&2\kk(m+1,n)\cdot J(1,\cdt,m)J_{\xi}(m+1,\cdt,n)
\nonumber\\&-&2\kk(1,m)\cdot J(m+1,\cdt,n)
J_{\xi}(1,\cdt,m)\nonumber\\&+&(\kk(1,m)-\kk(m+1,n))_{\xi}\nonumber\\&\times &J(1,\cdt,m)\cdot J(m+1,\cdt,n)
\eea
\noindent and
\bea
\{J(\alpha),J(\beta),J(\gamma)\}_{\xi}&=&J(\alpha)\cdot(J(\gamma)J_{\xi}(\beta)-J(\beta)
J_{\xi}(\gamma))\nonumber\\&-&J(\gamma)\cdot(J(\beta)J_{\xi}(\alpha)
-J(\alpha)J_{\xi}(\beta))
\eea
\noindent in which we used the compact notation $\alpha=(1,\cdt,m)$, $\beta=(m+1,\cdt,k)$ and $\gamma=(k+1,\cdt,n)$.
We mention that the current $J_\xi$ is the same of \cite{giele} appearing in the gluonic recursion relations.
This current satisfy the identities
\bea
(a)&& \kappa(1,n)\cdot J(1,\cdots,n)=0\nonumber\\
(b)&& \ J(1,\cdots,n)=(-)^{n-1}J(n,\cdots,1)\nonumber\\
(c)&& \sum_{Cyc(1,\cdots,n)}J(1,\cdots,n)=0 \label{97} \eea
\noindent The
recursion relation include of clock-wise orientation for the
labels $1,2,\cdots ,n$ that is number of all diagrams occurring in
recursion relation; for 3-photon is 3, and for 4-photon is 10; Fig.~\ref{photons} and
Fig.~\ref{feynman}.

\section{From Currents To Amplitudes And \\ Cross Sections}
Once one is given by the currents, the amplitudes, demanding
overall momentum conservation $\kk^{\m}(1,n)=0$, are given by \bea
\mathbf{M}(1,\cdots ,n)&=&\imath\ \widehat{J}^{\mu}(1,\cdots ,n-1)
\times\widehat{J}_{\mu}(n)\ \kappa^{2}(1,n-1)\mid_{\kappa(1,n)=0}
\nonumber\\&=&i(-)^{n-1}(ie)^{n-2} \sum_{P(1,\cdots
,n-1)}C(1\cdots n-1;+\frac{n-4}{2}\pi)\nonumber\\&& \times
J(1,\cdots ,n-1)\cdot
J(n)\kappa^{2}(1,n-1)\mid_{\kappa(1,n)=0}\nonumber\\&=&
(-1)^{n-1}i^{n}e^{n-2} \sum_{P(1,\cdots ,n-1)}C(1\cdots
n-1;+\frac{n-4}{2}\pi) \varepsilon(1,\cdots ,n) \label{8}
\nonumber\\&&\eea \noindent
where $\varepsilon$-function is given
by \bea \varepsilon(1,\cdots ,n)=J(1,\cdots ,n-1)\cdot
J(n)\kappa^{2}(1,n-1)\mid_{\kappa(1,n)=0} \label{9} \eea By the
properties we know about the currents, we can show the following
for $\varepsilon(1,\cdots ,n)$:
\begin{enumerate}
\item $\varepsilon$ is invariant under cyclic permutations,
\bea
\varepsilon(1,\cdots ,n)=\varepsilon(m+1,\cdots ,n,1,\cdots ,m)
\label{10} \eea \noindent
\item $\varepsilon$ has a reflective property,
\bea
\varepsilon(1,\cdots ,n)=(-)^{n}\varepsilon(n,\cdots ,1)
\eea
\item The sub-cyclic sum equals zero,
\bea
\sum_{Cyc(1,\cdots ,n-1)}\varepsilon(1,\cdots ,n)=0 \label{11}
\eea
\item The quantity $\varepsilon$ is gauge invariant.
\end{enumerate}
These properties are proven in \cite{giele}. For
the cross section one must square the amplitude eq.(\ref{8}), yielding
\bea
\mid \mathbf{M}(1,\cdots ,n)\mid^{2}&=& (e)^{2n-4}\sum_{P(1,\cdots
,n-1)}C(1\cdots n-1;+\frac{n-4}{2}\pi) \varepsilon(1,\cdots
,n)\nonumber\\&&\times \sum_{P(1,\cdots ,n-1)}C(1\cdots n-1;+\frac{n-4}{2}\pi)
\varepsilon^{\star}(1,\cdots ,n)
\eea

\section{Solution Of The Recursion Relation For Special Helicity Configurations}
The cross sections often contain contributions from a large number of diagrams. The standard
methods of calculation, where an amplitude is squared and summed
over all polarizations often become unpractical.
For these reasons it became necessary to try other
methods of calculation. Most of these procedures calculate
amplitudes for specific polarization states of the particles, for
which usually helicities are taken. Since at the high energies considered most of the
fermion masses can be neglected, the description in terms of
helicity states is easily incorporated by using $1\pm\gamma^{5}$
projections of Dirac spinors. The covariant description of the
photon (gluon) helicity in terms of a suitable polarization vector
$\epsilon$ is also needed.
The recursion relation can be used to calculate
step by step any current with a certain number of photons having a
specific helicity configuration. For some special helicity
configurations the results become so simple that a generalization
to an arbitrary number of photons presents itself. The helicity
configurations for which this is possible are those in which all
photon helicities are the same or all but one are the same. In
this way we find $J(1+,2+,\cdots ,n+)$  and $J(1-,2+,3+,\cdots ,n+)$.

The photonic recursion relation will be solved for two state
helicity. The configurations are those where all helicities are
the same or all but one are the same.
The inner product between 4-vectors has been replaced by spinor
contractions, one can choose a gauge
for the helicity spinors such that for all polarization vectors
$\epsilon_{i}(q)\cdot\epsilon_{j}(q)=0$ ($q$ is gauge spinor and for more, see Appendix). Through the
recursion relation the currents keep this orthogonality property.
Thus the 4-vertex contributions vanish, \cite{giele, str}. So, the remained purely
photonic current is
\bea
J_{\m}(1,\cdots ,n)&=&\frac{1}{\kappa^{2}(1,n)}\sum^{n-1}_{m=1}
[J(1,\cdots ,m),J(m+1,\cdots ,n)]_{\m}\label{3}
\eea
\noindent
We can choose following forms polarization state for $J(1+,2+,\cdots ,n+)$
\bea
\e^{+}_{\m}(i,q)=\frac{\langle q^{-}\mid\gamma_{\m}\mid k_{i}^{-} \rangle }
{\sqrt{2}\langle qk_{i}\rangle }\ ,\ \ \ \ 1\leq i\leq n
\label{19}
\eea
\noindent Where $q^{\mu}$ is some four-vector different from $k$
and for $J(1-,2+,\cdots ,n+)$
\bea
\e^{-}_{\m}(1,k_{2})=-\frac{\langle k^{+}_{2}\mid\gamma_{\m}\mid k^{+}_{1}\rangle }
{\sqrt{2}[k_{2}k_{1}]},\;\;\;\;\;
\e^{+}_{\m}(i,k_{1})=\frac{\langle k^{-}_{1}\mid\gamma_{\m}\mid k^{-}_{i}\rangle }
{\sqrt{2}\langle k_{1}k_{i}\rangle}\ ,\ \ \ \ 2\leq i\leq n \label{20}
\eea
\noindent where $k_{2}$ in $\e^{-}(1,k_{2})$ is as like as $q$ in $\e^{+}_{\m}(i,q)$,
represents the gauge spinor. The recursion relation takes the simple form
\bea
J_{\m}(1,\cdots ,n)&=&\frac{1}{\kappa^{2}(1,n)}\sum^{n-1}_{m=1}
\big(\kappa(m+1,n)\cdot J(1,\cdots ,m)J_{\m}(m+1,\cdots ,n)\nonumber\\&&
-\kappa(1,m)\cdot J(m+1,\cdots ,n)J_{\m}(1,\cdots ,m)\big)\label{4}
\eea
In Appendix we show
\bea
&&J_{\m}(1+,2+,\cdots ,m+)=\frac{\langle  \kk(1,m)q\rangle \langle q^{-}\mid\gamma_{\m}\mid
\kk^{-}(1,m)\rangle }
{2^{\frac{m}{2}}\langle\langle q1,mq\rangle\rangle}
\label{5}
\eea
\noindent where $\langle\langle q1,mq\rangle\rangle=\langle q1\rangle\langle12\rangle\cdots
\langle m-1,m\rangle\langle mq\rangle$. These equations reduce to
eq.(\ref{19}) for one photon. The general case is proven by  induction (see Appendix).
To evaluate of $J(1-,2+,\cdots ,n+)$, we have
\bea
J_{\m}(1-)=\e^{-}_{\m}(1,k_{2})=-\frac{\langle 2^{+}\mid\gamma_{\m}\mid 1^{+}\rangle }
{\sqrt{2}[21]}
\eea
\noindent and we can show that $J_{\m}(1-,2+)=0$, So we can show
\bea
J_{\m}(1-2+\cdots n+)=-2^{-\frac{1}{2}}J_{\m}(2+\cdots n+)\frac{\langle
1n\rangle [n\kk(1,n)]\langle  1\kk(1,n)\rangle }{\kappa^{2}(1,n-1)\kappa^{2}(1,n)}
\label{111}
\eea
This is proven in \cite{giele}. Complex conjugation of the currents eq.(\ref{5}) and eq.(\ref{111}) gives
$J(1-2-\cdots n-)$ and $J(1+2-\cdots n-)$.

Since we have solved the recursion relation for currents in cases
of specific helicity configurations we can calculate amplitudes
for these situations as well. We do this for n-photon scattering
with and without the production of other particles.

From the currents we make $\varepsilon$-functions and from them
the helicity amplitudes according to eq.(\ref{8}) and
eq.(\ref{9}). With the explicit expression for $J^{\mu}(2+\cdots n+)$
in eq.(\ref{5}) we have
\bea
\varepsilon(1\pm2+\cdots n+)&=&\kappa^{2}(2,n)\e^{\pm}_{\m}(1,2)
J^{\m}(2+\cdots n+)\mid_{\kappa(1,n)=0}\nonumber\\&=&
\kappa^{2}(2,n)\e^{\pm}_{\m}(1,2)\frac{\langle  \kk(1,m)q\rangle \langle q^{-}\mid\gamma_{\m}\mid \kk^{-}(1,m)\rangle }
{2^{\frac{m}{2}}\langle\langle q1,mq\rangle\rangle}\Big|_{\kappa(1,n)=0}=0\nonumber\\&&
\eea
The vanishing of this $\varepsilon$-function is due to the
overall momentum conservation, which leads to a vanishing
$\kappa^{2}(2,n)$ (because, $\kappa(1,n)=0\rightarrow\kappa(2,n)=-k_{1}
\rightarrow\kappa^{2}(2,n)=k^{2}_{1}=0$). With the cyclic symmetry of the
$\varepsilon$-function, also the $\varepsilon$-function with one
negative helicity in an arbitrary position vanishes. The helicity
amplitude then vanishes as well
\bea
\textbf{M}(1\pm2+\cdots n+)=0
\eea
The first non-trivial helicity amplitude is $\textbf{M}(1-2-3+\cdots n+)$, for which we have
\bea
\varepsilon(1-2-3+\cdots n+)&=&\kk^{2}(2,n)\e^{-}_{\m}(1,3)J^{\m}(2-3+\cdots n+)
\Big|_{\kk(1,n)=0}
\eea
This is not vanishing for conserved momentum, because $J^{\m}(2-3+\cdots n+)$ has
$\frac{1}{\kk^{2}(2,n)}$ so becomes
\bea
\varepsilon(1-2-3+\cdots n+)&=&
(-\frac{\langle 3^{+}\mid\gamma_{\m}\mid1^{+}\rangle }{\sqrt{2}[31]})
J^{\m}(3+\cdots n+)\nonumber\\
&&(-2^{-\frac{1}{2}}\frac{\langle
2\ n\rangle [n\ \kk(2,n)]\langle  2\ \kk(2,n)\rangle }{\kappa^{2}(2,n-1)})\Big|_{\kk(1,n)=0}
\nonumber\\&=&
(-\frac{\langle 3^{+}\mid\gamma_{\m}\mid1^{+}\rangle }{\sqrt{2}[31]})
(\frac{\langle  \kk(3,n)\ 2\rangle \langle 2^{-}\mid\gamma_{\m}\mid \kk^{-}(3,n)\rangle }
{2^{\frac{n-3}{2}}\langle\langle 23,n2\rangle\rangle})
\nonumber\\&&\times(-2^{-\frac{1}{2}}\frac{\langle
2\ n\rangle [n\ \kk(2,n)]\langle  2\ \kk(2,n)\rangle }{\kappa^{2}(2,n-1)})\Big|_{\kk(1,n)=0}
\nonumber\\&=&
2^{(1-n)/2}\frac{\langle12\rangle^{4}}{\langle\langle12,n1\rangle\rangle}
\eea
In the general case and from the recursion relation, we find \cite{giele}
\bea
&&\varepsilon(1-3+\cdots m+2-(m+1)+\cdots n+)\nonumber\\&&=
2^{(1-n)/2}\frac{\langle12\rangle^{4}}
{\langle13\rangle\langle34\rangle\cdots \langle
m2\rangle\langle2(m+1)\rangle\cdots \langle n1\rangle}
\eea
which gives the amplitude
\bea
&&\mathbf{M}(1-,2-,\cdots ,n+)\nonumber\\&&\ \ \ \ =
(-)^{n-1}i^{n}e^{n-2}2^{\frac{1-n}{2}}\langle12\rangle^{4} \sum_{P(1,\cdots ,n-1)}C(1\cdots
n-1;+\frac{n-4}{2}\pi)\frac{1}{\langle\langle12,n1\rangle\rangle}
\nonumber\\&&
\eea
We mention that the recursion relation for $n$-photon scattering
process in U(1) noncommutative space, is very similar to recursion relation for $n$-gluon scattering
case in SU(3) on ordinary space. By this, we drive photons process scattering
amplitudes for special cases in helicity. The main observation is that the
helicity configuration is responsible factor for non-vanishing of scattering

\subsection{MHV Amplitudes For Fermion-Photon Interactions Of NCQED}
We derive recursion relation for electron and positron of NCQED in \cite{jafari2}.
We can write electron-photon current in following equation
\bea
\widehat{J}(Q,1,\cdots ,n)=e^{n}\sum_{P(1,\cdots ,n)}\exp(i\kk(1,n)Q;-i\kk(1,n);)
J(Q,1,\cdots ,n)
\label{14}
\eea
\noindent where
\bea
J(Q,1,\cdots ,n)=\frac{-1}{Q\sl+\kk\sl(1,n)-m_0}
\sum_{m=0}^{n-1}h(n,m)J(Q,1,\cdots ,m)\jsl(m+1,\cdots ,n)
\label{16} 
\eea
\noindent in which
\bea
h(n,m)=\frac{1}{2}(-1)^{n-m}(i)^{n-m-1}
(\,{\rm e}^{-i\frac{n-m-3}{2}\pi}
+(-1)^{n-m-1}\,{\rm e}^{i\frac{n-m-3}{2}\pi})  \label{17}
\eea
with $Q$ and $m_{0}$ as electron's momentum and mass respectively. The fermion
case in NCQED is different by quark case in QCD, for reason, we must evaluate helicity
amplitudes step by step.

We know electrons obey free Dirac equation and these have
ordinary operator projections, so in following calculations we
keep same gauge spinor ($Q$- electron's momentum) for all photons and we define
$\kk(1,i;Q)=Q+\kk(1,i)$ and $Qi=Q+k_{i}$ also $\kk\sl(1,i;Q)=Q\sl+\sum_{l=1}^{i} k_{l}\sl$ and
$\mm\kk^{\pm}(1,i;Q)\rangle\langle\kk^{\pm}(1,i;Q)\mm=\mm Q^{\pm}\rangle\langle
Q^{\pm}\mm+\mm\kk^{\pm}(1,i)\rangle\langle\kk^{\pm}(1,i)\mm$, in spinorial language
\bea
\widehat{J}(Q+)=\langle Q^{+}\!\mm
\eea
and
\bea
\widehat{J}(Q+,1+)&=&-\frac{1}{\kk^{2}(1;Q)}\langle
Q^{+}\mm\esl^{+}(1,Q)\kk\sl(1;Q)\nonumber\\&=&
-\frac{1}{\kk^{2}(1;Q)}\langle Q^{+}\mm (+\frac{\langle Q^{-}\mm\gamma
^{\m}\mm 1^{-} \rangle }{\sqrt{2}\langle Q1 \rangle
}\gamma_{\m})(\sum_{\lambda=\pm}\mm \kk^{\lambda}(1;Q) \rangle \langle
\kk^{\lambda}(1;Q)\!\!\mm)\nonumber\\&& \eea
One mentions
\bea
\langle Q^{\lambda_{1}}\mm\gamma ^{\m}\mm 1^{\lambda_{2}}
\rangle \gamma_{\m}=2\mm 1^{\lambda_{2}} \rangle \langle
Q^{\lambda_{1}}\mm+ 2\mm Q^{-\lambda_{1}} \rangle \langle
1^{\lambda_{2}}\mm
\eea
and so
\bea
\widehat{J}(Q+,1+)=-\sqrt{2}\frac{\langle Q\ \kk(1;Q) \rangle }{\langle
1Q \rangle \langle Q1 \rangle }\langle \kk(1;Q)^{+}\mm
\eea
For the next case we have
\bea
\widehat{J}(Q+,1+,2+)&=&-\frac{1}{\kk^{2}(1,2;Q)}(h(2,0)J(Q+)\jsl^{+}(1+,2+)
+h(2,1)J(Q+,1+)\nonumber\\&\times &\esl^{+}(2,Q))\kk\sl(1,2;Q)
=-\frac{1}{(q2)^{2}}(h(2,0)A+h(2,1)B)
\eea
where
\bea
A&=&\langle Q^{+}\mm (\langle \kk(1,2)\ Q \rangle \frac{\langle Q^{-}\mm\gamma
^{\m}\mm \kk^{-}(1,2) \rangle }{2\langle Q1 \rangle \langle 12
\rangle \langle 2Q \rangle }\gamma_{\m})\sum_{\lambda=\pm}\mm \kk^{\lambda}(1,2;Q) \rangle \langle \kk^{\lambda}(1,2;Q)\mm
\nonumber\\&=&
\frac{\langle \kk(1,2)\ Q \rangle [Q\ \kk(1,2)]\langle Q\ \kk(1,2;Q)
\rangle }{\langle Q1 \rangle \langle 12 \rangle \langle 2Q \rangle
}\langle \kk^{+}(1,2;Q)\mm
\eea
and
\bea B&=&(-\frac{\sqrt{2}\langle Q\ \kk(1;Q) \rangle
}{\langle 1Q \rangle \langle Q1 \rangle }\langle \kk^{+}(1;Q)\!\!\mm)
(\frac{\langle Q^{-}\!\!\mm\gamma ^{\m}\!\!\mm 1^{-} \rangle
}{\sqrt{2}\langle Q1 \rangle }\gamma_{\m})\sum_{\lambda=\pm}\mm \kk^{\lambda}(1,2;Q)
\rangle \langle \kk^{\lambda}(1,2;Q)\!\!\mm \nonumber\\&=&
-2\frac{\langle Q\ \kk(1;Q) \rangle [\kk(1;Q)\ 2]\langle Q\ \kk(1,2;Q) \rangle
}{\langle Q1 \rangle \langle 1Q \rangle \langle Q2 \rangle
}\langle \kk^{+}(1,2;Q)\!\!\mm
\eea
Noting
\bea
\langle Q\ \kk(1;Q) \rangle [\kk(1;Q)\
2]=\langle Q^{-}\ \kk^{+}(1;Q) \rangle \langle \kk^{+}(1;Q)\ 2^{-} \rangle
=-\langle Q^{-}\mm 2\sl\mm 2^{-} \rangle
\eea
which from momentum conservation we have $\kk(1,2;Q)=0\rightarrow \kk(1;Q)=-2$ so
\bea
\langle Q\ \kk(1;Q) \rangle [\kk(1;Q)\ 2]=-\langle Q^{-}\mm 2\sl\mm 2^{-}\rangle =0
\eea
giving B=0. So we get
\bea
\widehat{J}(Q+,1+,2+)&=&-\frac{1}{\kk^{2}(1,2;Q)}h(2,0)A=\frac{\langle
Q\ \kk(1,2;Q) \rangle }{\langle 1Q \rangle \langle 12 \rangle \langle 2Q
\rangle }\langle \kk^{+}(1,2;Q)\mm
\eea
The above can be generalized to
\bea
\widehat{J}(Q+,1+,2+,\cdots,n+)&=&(-1)^{n}2^{\frac{2-n}{2}}\frac{\langle
Q\ \kk(1,n;Q) \rangle }{\langle \langle 1Q,nQ \rangle  \rangle }\langle
\kk^{+}(1,n;Q)\mm
\eea
We prove it by induction. It is correct for $m< n$ so
\bea
&&\widehat{J}(Q+,1+,2+,\cdots,n+)\nonumber\\&&=-\frac{1}{\kk^{2}(1,n;Q)}\sum_{m=0}^{n-1}
h(n,m)J(Q+,1+,\cdots,m+)\jsl^{+}((m+1)+,\cdots,n+)\kk\sl(1,n;Q)\nonumber\\&&
=-\frac{1}{\kk^{2}(1,n;Q)}\sum_{m=0}^{n-1} h(n,m)(-1)^{m}2^{\frac{m-2}{2}}
\frac{\langle Q\ \kk(1,m;Q) \rangle }{\langle \langle 1Q,mQ \rangle
 \rangle }\langle \kk^{+}(1,m;Q)\mm \nonumber\\&&\times\frac{\langle
\kk(m+1,n)Q\rangle \langle Q^{-}\mid\gamma_{\m}\mid
\kk^{-}(m+1,n)\rangle } {2^{\frac{n-m}{2}}\langle\langle
Q(m+1),nQ\rangle\rangle}\gamma^{\m}(\sum_{\lambda=\pm}\mm
\kk^{\lambda}(1,n;Q) \rangle \langle \kk^{\lambda}(1,n;Q)\mm )\nonumber\\&&
=-\frac{1}{\kk^{2}(1,n;Q)}\sum_{m=0}^{n-1} h(n,m)(-1)^{m}2^{\frac{m-2}{2}}
\frac{\langle \kk(m+1,n)Q\rangle\langle Q\ \kk(1,m;Q) \rangle }{\langle
\langle 1Q,mQ \rangle  \rangle } \nonumber\\&&[\kk(1,m;Q)\
\kk(m+1,n)]\langle Q^{-}\mm\times\frac{2}
{2^{\frac{n-m}{2}}\langle\langle
Q(m+1),nQ\rangle\rangle}\nonumber\\&&\times(\sum_{\lambda=\pm}\mm
\kk^{\lambda}(1,n;Q)\rangle \langle \kk^{\lambda}(1,n;Q)\mm )\nonumber\\&&
\eea
but
\bea
&& \langle Q\ \kk(1,m;Q)\rangle [\kk(1,m;Q)\ \kk(m+1,n)]=\langle Q^{-}\ \kk^{+}(1,m;Q) \rangle
\nonumber\\&&\times\langle
\kk^{+}(1,m;Q)\ \kk^{-}(m+1,n) \rangle = \langle Q^{-}\mm
\kk\sl(1,m;Q)\mm \kk^{-}(m+1,n) \rangle
\eea
which from momentum conservation we have $\kk(1,m;Q)+\kk(m+1,n)=0\rightarrow
\kk(1,m;Q)=-\kk(m+1,n)$. So all above terms vanish for $m=1$ to $n-1$, but
for $m=0$ we have
\bea
&&\widehat{J}(Q+,1+,2+,\cdots,n+)\nonumber\\&&=
-\frac{1}{\kk^{2}(1,n;Q)} h(n,0)\langle Q^{+}\mm (\sum_{\lambda=\pm}\mm \kk^{\lambda}(1,n;Q) \rangle
\langle \kk^{\lambda}(1,n;Q)\mm )
\nonumber\\&&\times(\frac{\langle \kk(1,n)Q\rangle \langle
Q^{-}\mid\gamma_{\m}\mid \kk^{-}(1,n)\rangle }
{2^{\frac{n}{2}}\langle\langle
Q1,nQ\rangle\rangle}\gamma^{\m})\nonumber\\&& =-\frac{1}{\kk^{2}(1,n;Q)}
h(n,0)[Q\ \kk(1,n)]\langle Q\ \kk(1,n;Q)\rangle \langle  \kk(1,n)Q\rangle
\frac{\langle \kk^{+}(1,n;Q)\mm} {2^{\frac{n-2}{2}}\langle\langle
Q1,nQ\rangle\rangle}\nonumber\\&& =-\frac{h(n,0)\langle  Q\
 \kk(1,n;Q)\rangle} {2^{\frac{n-2}{2}}\langle\langle
Q1,nQ\rangle\rangle}\langle \kk^{+}(1,n;Q)\mm\nonumber\\&&
\eea
in which
\bea
-h(n,0)2^{\frac{2-n}{2}}&=&2^{\frac{-n}{2}}(-1)^{n+1}i^{n-1}(e^{-i\frac{n-3}{2}\pi}
+(-1)^{n-1}e^{i\frac{n-3}{2}\pi}))\nonumber\\&&
\eea
For any $n$ (odd or even), we have $i^{n-1}(e^{-i\frac{n-3}{2}\pi} +(-1)^{n-1}e^{i\frac{n-3}{2}\pi})=-2$
so
$$
-h(n,0)2^{\frac{2-n}{2}}=(-1)^{n}2^{\frac{2-n}{2}}
$$
This completes the proof. For
\bea
\widehat{J}(Q+,1-)&=&-\frac{1}{\kk^{2}(1;Q)}\langle
Q^{+}\mm\esl^{-}(1,Q)\kk\sl(1;Q)\nonumber\\&=&
-\frac{1}{\kk^{2}(1;Q)}\langle Q^{+}\mm (-\frac{\langle Q^{+}\mm\gamma
^{\m}\mm 1^{+} \rangle }{\sqrt{2}[Q1]}\gamma_{\m})(\sum_{\lambda=\pm}\mm \kk^{\lambda}(1;Q) \rangle \langle
\kk^{\lambda}(1;Q)\mm)\nonumber\\&&
=-\frac{1}{\kk^{2}(1;Q)}\langle Q^{+}\mm (-\frac{
2\mm1^{+} \rangle \langle Q^{+}\mm+2\mm Q^{-} \rangle \langle
1^{-}\mm}{\sqrt{2}[Q1]})\nonumber\\&& \times(\sum_{\lambda=\pm}\mm \kk^{\lambda}(1;Q) \rangle \langle
\kk^{\lambda}(1;Q)\mm)\nonumber\\&=&
\sqrt{2}\frac{1}{\kk^{2}(1;Q)}[Q\ \kk(1;Q)] \langle \kk(1;Q)\ 1 \rangle
\frac{1}{[Q1]}\nonumber\\&=& \sqrt{2}\frac{1}{\kk^{2}(1;Q)}\langle
Q^{+}\mm \kk\sl(1;Q)\mm 1^{-} \rangle \frac{1}{[Q1]}\nonumber\\&&=0
\eea
and
\bea
\widehat{J}(Q+,1-,2+)&=&-\frac{1}{\kk^{2}(1,2;Q)}\big(h(2,0)J(Q+)\jsl(1-,2+)
+h(2,1)J(Q+,1-)\esl^{+}(2,Q)\nonumber\\&+&h(2,1)J(Q+,2+)\esl^{-}(1,Q)\big)\kk\sl(1,2;Q)
\nonumber\\&=&
-\frac{1}{\kk^{2}(1,2;Q)}(h(2,0)A+h(2,1)(B+C))\nonumber\\&&
\eea
But, $B$ and $A$ vanish and from (\ref{111}) we get to
\bea
\widehat{J}(Q+,1-,2+)&=&-\frac{h(2,1)C}{\kk^{2}(1,2;Q)}
\eea
where
\bea
C&=&(\frac{\langle Q\ \kk(2;Q) \rangle }{\langle Q1 \rangle \langle 12
\rangle \langle 2Q \rangle }\langle \kk^{+}(2;Q)\mm)\nonumber\\&\times &
(-\frac{\langle Q^{+}\mm\gamma^{\m}\mm 1^{+} \rangle
}{\sqrt{2}[Q1]}\gamma_{\m})(\sum_{\lambda=\pm}\mm \kk^{\lambda}(1,2;Q) \rangle \langle \kk^{\lambda}(1,2;Q)\mm)
\nonumber\\&=&-2\frac{\langle Q\ \kk(2;Q)
\rangle [\kk(2;Q)\ Q]\langle 1\ \kk(1,2;Q) \rangle }{\sqrt{2}[Q1]\langle Q1
\rangle \langle 12 \rangle \langle 2Q \rangle }\langle \kk^{+}(1,2;Q)\mm
\eea
One mentions
\bea
\langle Q\ \kk(2;Q)\rangle [\kk(2;Q)\ Q]&=&\langle Q^{-}\ \kk^{+}(2;Q)
\rangle \langle \kk^{+}(2;Q)\ Q^{-} \rangle \nonumber\\
=\langle Q^{-}\mm \kk\sl(2;Q)\mm
Q^{+} \rangle = \langle Q^{-}\mm -1\sl\mm Q^{+} \rangle &=&-\langle
Q\ 1 \rangle [1\ Q]
\eea
so
\bea
\widehat{J}(Q+,1-,2+)&=&\sqrt{2}\frac{h(2,1)}{\kk^{2}(1,2;Q)}
\frac{\langle Q\ 1 \rangle [1\ Q]\langle 1\ \kk(1,2;Q) \rangle
}{[Q\ 1]\langle Q1 \rangle \langle 12 \rangle \langle 2Q \rangle
}\langle \kk^{+}(1,2;Q)\mm \nonumber\\&=&- \sqrt{2}\frac{h(2,1)}{\kk^{2}(1,2;Q)}
\frac{\langle Q\ 1 \rangle \langle 1\ \kk(1,2;Q) \rangle }{\langle \langle
Q1,2Q \rangle  \rangle }\langle \kk^{+}(1,2;Q)\mm \nonumber\\&&
\eea
in which h(2,1)=1. So we have
\bea
\widehat{J}(Q+,1-,2+)&=&\frac{-\sqrt{2}}{\kk^{2}(1,2;Q)} \frac{\langle
Q\ 1 \rangle \langle 1\ \kk(1,2;Q)\rangle }{\langle \langle Q1,2Q \rangle
\rangle }\langle \kk^{+}(1,2;Q)\mm \nonumber\\&&
\eea
It is generalized to
\bea
\widehat{J}(Q+,1-,2+,\cdots,n+)&=&\frac{(-1)^{n-1}2^{\frac{n-1}{2}}}{\kk^{2}(1,n;Q)}
\frac{\langle Q\ 1 \rangle \langle 1\ \kk(1,n;Q)\rangle }{\langle \langle
Q1,nQ \rangle  \rangle }\langle \kk^{+}(1,n;Q)\mm \nonumber\\&&
\eea
We mention that despite the difference between the recursion relations
of NCQED and QCD cases, the helicity amplitude of special configuration
(all photons and fermion have positive helicity) we approach to same result.
Based on this observation, the generalization we suggest
seems reasonable, for which we shall present the proof in a forthcoming paper.

\appendix
\section{Spinor Formalism}
The spinor helicity formalism for massless vector bosons
\cite{kleiss, zhanf, kunst} is largely responsible for the
existence of extremely compact representations of tree and loop
partial amplitudes in QCD. It introduces a new set of kinematic
objects, spinor products, which neatly capture the collinear
behavior of these amplitudes. A (small) price to pay is that
automated simplification of large expressions containing these
objects is not always straightforward, because they obey nonlinear
identities. Here we will review the spinor helicity
formalism and some of the key identities. We begin with massless
fermions. Positive and negative energy solutions of the massless
Dirac equation are identical up to normalization conventions. One
way to see this is to note that the positive and negative energy
projection operators, $\Lambda_{+}(k)\sim u(k).\bar{u}(k)$ and
$\Lambda_{-}(k)\sim v(k).\bar{v}(k)$, are both proportional to
$k\sl$ in the massless limit. Thus the solutions of definite
helicity, $u_{\pm}(k) = \frac{1}{2}(1\pm \gamma_{5})u(k)$ and
$v_{\pm}(k)=\frac{1}{2}(1\pm\gamma_{5})v(k)$, can be chosen to be
equal to each other. (For negative energy solutions, the helicity
is the negative of the chirality or $\gamma_{5}$ eigenvalue.) A
similar relation holds between the conjugate spinors
$\bar{u}_{\pm}(k) = \frac{1}{2}\bar{u}(k)(1\pm \gamma_{5})$ and
$\bar{v}_{\pm}(k)=\frac{1}{2}\bar{v}(k)(1\pm\gamma_{5})$. Since we
will be interested in amplitudes with a large number of momenta,
and we use the shorthand notation $\mid i^{\pm} \rangle =\mid
k_{i}^{\pm} \rangle =u_{\pm}(k_{i})=v_{\mp}(k_{i})$ and $\langle
i^{\pm}\mid=\langle  k_{i}^{\pm}
\mid=\bar{u}_{\pm}(k_{i})=\bar{v}_{\mp}(k_{i})$ and we define the
basic spinor products by $\langle  ij \rangle =\langle  i^{-}\mid
j^{+} \rangle =\bar{u}_{-}(k_{i})u_{+}(k_{j})$ and $[ij]=\langle
i^{+}\mid j^{-} \rangle =\bar{u}_{+}(k_{i})u_{-}(k_{j})$. The
helicity projection implies that products like $\langle
i^{+}|j^{+}\rangle $ vanish. For numerical evaluation of the
spinor products, it is useful to have explicit formulae for them,
for some representation of the Dirac $\gamma$ matrices. In the
Dirac representation,
\bea \gamma^{0}= \left(
\begin{array}{cc}
1&0\\
0&-1\\
\end{array}
\right)\ \ \ \ \gamma^{i}= \left(
\begin{array}{cc}
0&\sigma^{i}\\
-\sigma^{i}&0\\
\end{array}
\right)\ \ \ \ \gamma^{5}= \left(
\begin{array}{cc}
0&1\\
1&0\\
\end{array}
\right)
\eea
The spinor products are, up to a phase, square roots
of Lorentz products. We'll see that the collinear limits of
massless gauge amplitudes have this kind of square-root
singularity, which explains why spinor products lead to very
compact analytic representations of gauge amplitudes, as well as
improved numerical stability. We would like the spinor products to
have simple properties under crossing symmetry, i.e. as energies
become negative. We define $[ij]$ through the identity \cite{dixon}
\bea
\langle  ij \rangle  [ji]= \langle  i^{-}
j^{+}\rangle \langle j^{+} i^{-}\rangle
=tr(\frac{1}{2}(1-\gamma_{5})k\sl_{i}k\sl_{j})=2k_{i}\cdot
k_{j}=s_{ij}
\eea
We also have the useful identities\\
- Gordon identity and projection operator:
\bea
\langle
i^{\pm}\mid\gamma^{\mu}\mid i^{\pm}\rangle =2k_{i}^{\mu}\ \ \ \
\mid i^{\pm}\rangle \langle
i^{\pm}\mid=\frac{1}{2}(1\pm\gamma_{5})k\sl_{i}
\eea
- Antisymmetry:
\bea
\langle  ij \rangle =- \langle ji\rangle \ \ \ \
[ij]=-[ji]\ \ \ \ \langle  ij \rangle =[ij]=0
\eea
- Fierz rearrangement
\bea
\langle  i^{+}\mid\gamma_{\mu}\mid j^{+}\rangle
\langle  k^{+}\mid\gamma^{\mu}\mid l^{+}\rangle =2[ik]\langle
lj\rangle
\eea
- Charge conjugation of current
\bea
\langle
i^{+}|\gamma^{\mu}|j^{+}\rangle  = \langle
j^{-}|\gamma^{\mu}|i^{-}\rangle
\eea
- Schouten identity
\bea
\langle  ij \rangle \langle  kl \rangle =\langle  ik \rangle
\langle  jl \rangle +\langle  il \rangle \langle  kj \rangle
\eea
In an $n$-point amplitude, momentum conservation, $\sum_{i=1}^{n}k^{\mu}_{i}=0$, provides one more identity
\bea
\sum_{i=1,i\neq j,k}^{n}[ji]\langle  ik \rangle =0
\eea
For a massless photon and with momentum $k^{\mu}$ we have in the axial gauge
\bea
\sum_{polarization}\epsilon^{\mu}\epsilon^{\star\nu}=-g^{\mu\nu}+\frac{(q^{\mu}k^{\nu}+k^{\mu}q^{\nu})}
{q\cdot k}
\eea
\noindent where $q^{\mu}$ is some four-vector different from $k^{\mu}$. The two degrees of freedom in the spin-1
field will be described with the usual right and left-oriented helicity vectors \cite{dixon}. These helicity vectors
$\epsilon_{\pm}$ have the following properties \cite{giele, kleiss, dixon}
\bea
\epsilon_{\pm}\cdot\epsilon^{\star}_{\mp}=0\;\;\;\;\;\;\;\;\;\;
k\cdot\epsilon_{\pm}=0\ \ \ \ q\cdot\epsilon_{\pm}=0
\nonumber\\
\epsilon_{\pm}^{\dag}=\epsilon_{\mp}\;\;\;\;\;\;\;\;\;\;
\epsilon_{\pm}\cdot\epsilon_{\mp}=-1
\eea
By these four properties in the next step, is to introduce a
spinor representation for the polarization vector for a massless
gauge boson of definite helicity ±1,
\bea
\epsilon^{\mu}_{\pm}(k,q)=\pm\frac{\langle
q^{\mp}\mid\gamma^{\mu}\mid k^{\mp}\rangle }{\sqrt{2}\langle
q^{\mp} \mid k^{\pm}\rangle }
\eea
\noindent This obeys all properties relations and the new spinor $q$ is called the gauge
spinor and may be chosen any spinor except for $k$. Making another
choice of the gauge spinor is a gauge transformation of the spin-1 field.

\section{Photonic Process}
In spinor formalism, 4-vertex vanishes, because the polarization vector has same gauge spinor, so
\bea
\e^{+}(i,q)\cdot\e^{+}(j,q)\sim \langle  q^{-}\mid\gamma_{\m}\mid i^{-}\rangle .\langle  q^{-}\mid\gamma^{\m}\mid j^{-}\rangle
=2[ij]\langle qq\rangle =0
\eea
Consider the two equal helicity case ($q$ is gauge spinor)
\bea
J_{\m}(1+,2+)&=&\frac{1}{\kappa^{2}(1,2)}(\frac{1}{2}\langle
2^{\pm}\mid\gamma_{\n}\mid 2^{\pm} \rangle \frac{\langle
q^{-}\mid\gamma^{\n}\mid 1^{-}\rangle }{\sqrt{2}\langle q1\rangle
}\frac{\langle  q^{-}\mid\gamma_{\m}\mid 2^{-}\rangle }
{\sqrt{2}\langle q2\rangle }\nonumber\\&-& \frac{1}{2}\langle
1^{\pm}\mid\gamma_{\n}\mid 1^{\pm}\rangle \frac{\langle
q^{-}\mid\gamma^{\n}\mid 2^{-}\rangle }{\sqrt{2}\langle q2\rangle
}\frac{\langle  q^{-}\mid\gamma_{\m}\mid 1^{-}\rangle }
{\sqrt{2}\langle q1\rangle })\nonumber\\&&
\eea
but, $\kk^{2}(1,2)=2[21]\langle 12\rangle $ so
\bea
J_{\m}(1+,2+)&=&\frac{([21]\langle q2\rangle \langle  q^{-}\mid\gamma_{\m}\mid 2^{-}\rangle -
[12]\langle q1\rangle \langle  q^{-}\mid\gamma_{\m}\mid 1^{-}\rangle )}{2[21]\langle 12\rangle \langle q1\rangle \langle q2\rangle }
\nonumber\\&=&
\frac{-(\langle q2\rangle \langle  q^{-}\mid\gamma_{\m}\mid 2^{-}\rangle +
\langle q1\rangle \langle  q^{-}\mid\gamma_{\m}\mid 1^{-}\rangle )}{2\langle 12\rangle \langle q1\rangle \langle 2q\rangle }
\nonumber\\&=&
\frac{(\langle 2q\rangle \langle  q^{-}\mid\gamma_{\m}\mid 2^{-}\rangle +
\langle 1q\rangle \langle  q^{-}\mid\gamma_{\m}\mid 1^{-}\rangle )}{2\langle 12\rangle \langle q1\rangle \langle 2q\rangle }
\nonumber\\&=&
\frac{\langle  q^{-}\mid\gamma_{\m}\mid 2^{-}\rangle \langle 2q\rangle +
\langle  q^{-}\mid\gamma_{\m}\mid 1^{-}\rangle \langle 1q\rangle }
{2\langle 12\rangle \langle q1\rangle \langle 2q\rangle }\nonumber\\&=&
\frac{\langle  q^{-}\mid\gamma_{\m}\mid 2^{-}\rangle \langle 2^{-}q^{+}\rangle +
\langle  q^{-}\mid\gamma_{\m}\mid 1^{-}\rangle \langle 1^{-}q^{+}\rangle }
{2\langle 12\rangle \langle q1\rangle \langle 2q\rangle }\nonumber\\&=&
\frac{\langle  q^{-}\mid\gamma_{\m}2\sl\mid q^{+}\rangle +
\langle  q^{-}\mid\gamma_{\m}1\sl\mid q^{+}\rangle }
{2\langle 12\rangle \langle q1\rangle \langle 2q\rangle }\nonumber\\&=&
\frac{\langle \kk(1,2)q\rangle \langle  q^{-}\mid\gamma_{\m}\mid \kk^{-}(1,2)\rangle }
{2\langle q1\rangle \langle 12\rangle \langle 2q\rangle }\nonumber\\&&
\label{two}\eea
This is generalizable to
\bea
J_{\m}(1+,2+,\cdots ,m+)=\frac{\langle  \kk(1,m)q\rangle \langle q^{-}\mid\gamma_{\m}\mid
\kk^{-}(1,m)\rangle }
{2^{\frac{m}{2}}\langle\langle q1,mq\rangle\rangle}
\label{555}
\eea
\noindent where $\langle\langle q1,mq\rangle\rangle=\langle q1\rangle\langle12\rangle\cdots
\langle m-1,m\rangle\langle mq\rangle$. The equation reduces to
eq.(\ref{19}) for one photon. The general equation is proven by
induction; suppose eq.(\ref{555}) to be valid for $m< n$ then using
eq.(\ref{555}) we find
\bea
&& J_{\m}(1+,2+,\cdots ,n+)=\frac{1}{\kappa^{2}(1,n)}\sum^{n-1}_{m=1}\Big{(}\nonumber\\&&
\kappa_{\n}(m+1,n)\frac{\langle  \kk(1,m)q\rangle \langle q^{-}\mid\gamma^{\n}\mid \kk^{-}(1,m)\rangle }
{2^{\frac{m}{2}}\langle\langle q1,mq\rangle\rangle}
\frac{\langle  \kk(m+1,n)q\rangle \langle q^{-}\mid\gamma_{\m}\mid \kk^{-}(m+1,n)\rangle }
{2^{\frac{n-m}{2}}\langle\langle q(m+1),nq\rangle\rangle}
\nonumber\\&&- \kappa_{\n}(1,m)\frac{\langle  \kk(m+1,n)q\rangle \langle q^{-}\mid
\gamma^{\n}\mid \kk^{-}(m+1,n)\rangle }
{2^{\frac{n-m}{2}}\langle\langle q(m+1),nq\rangle\rangle}\frac{\langle  \kk(1,m)q\rangle \langle q^{-}\mid\gamma_{\m}\mid \kk^{-}(1,m)\rangle }
{2^{\frac{m}{2}}\langle\langle q1,mq\rangle\rangle}\Big{)}\nonumber\\&&\ \ \ \ \ \ \ \ \ \ \ \ \ \ \ \ \ \ \ \ \ \ \ \ \ \ \
=\frac{1}{\kappa^{2}(1,n)}\sum^{n-1}_{m=1}\frac{\mathbb{A}-\mathbb{B}}
{2^{\frac{n}{2}}\langle\langle q1,nq\rangle\rangle}
\eea
But, from Gordon identity
\bea &&
\langle  \kk(1,m)q\rangle \kk_{\n}(m+1,n)\langle q^{-}\mid\gamma^{\n}\mid \kk^{-}(1,m)\rangle \nonumber\\&&
=\langle  \kk(1,m)q\rangle \frac{1}{2}\langle \kk^{\pm}(m+1,n)\mid\gamma_{\n}\mid\kk^{\pm}(m+1,n)\rangle
\langle q^{-}\mid\gamma^{\n}\mid \kk^{-}(1,m)\rangle \nonumber\\&&
=\langle  \kk(1,m)q\rangle [\kk(m+1,n)\kk(1,m)]\langle q\kk(m+1,n)\rangle \nonumber\\&&
\eea
and by similar way
\bea &&
\kk_{\n}(1,m)\langle q^{-}\mid\gamma^{\n}\mid \kk^{-}(m+1,n)\rangle \nonumber\\&&
=\frac{1}{2}\langle \kk^{\pm}(1,m)\mid\gamma_{\n}\mid\kk^{\pm}(1,m)\rangle
\langle q^{-}\mid\gamma^{\n}\mid \kk^{-}(m+1,n)\rangle \nonumber\\&&
=[\kk(1,m)\kk(m+1,n)]\langle q\kk(1,m)\rangle \nonumber\\&&
\eea
so
\bea
\mathbb{A}&=&\langle  \kk(1,m)q\rangle [\kk(m+1,n)\kk(1,m)]\langle q\kk(m+1,n)\rangle
\nonumber\\&\times&
\langle  \kk(m+1,n)q\rangle \langle q^{-}\mid\gamma_{\m}\mid \kk^{-}(m+1,n)\rangle
\eea
and
\bea
\mathbb{B}&=&\langle  \kk(m+1,n)q\rangle [\kk(1,m)\kk(m+1,n)]\langle q\kk(1,m)\rangle
\nonumber\\&\times &
\langle  \kk(1,m)q\rangle \langle q^{-}\mid\gamma_{\m}\mid \kk^{-}(1,m)\rangle
\eea
when
\bea
\mathbb{A}-\mathbb{B}&=&\langle  \kk(1,m)q\rangle [\kk(m+1,n)\kk(1,m)]\langle q\kk(m+1,n)\rangle
\nonumber\\&\times &
\langle  \kk(m+1,n)q\rangle \langle q^{-}\mid\gamma_{\m}\mid \kk^{-}(m+1,n)\rangle
\nonumber\\&-&
(-\langle  q\kk(m+1,n)\rangle )(-[\kk(m+1,n)\kk(1,m)])(-\langle \kk(1,m)q\rangle )
\nonumber\\&\times &
\langle  \kk(1,m)q\rangle \langle q^{-}\mid\gamma_{\m}\mid \kk^{-}(1,m)\rangle
\nonumber\\&=&
\langle  \kk(1,m)q\rangle [\kk(m+1,n)\kk(1,m)]\langle q\kk(m+1,n)\rangle
\nonumber\\&\times &
(\langle  \kk(m+1,n)q\rangle \langle q^{-}\mid\gamma_{\m}\mid \kk^{-}(m+1,n)\rangle
+\nonumber\\&&
\langle  \kk(1,m)q\rangle \langle q^{-}\mid\gamma_{\m}\mid \kk^{-}(1,m)\rangle )\nonumber\\&&
\eea
Similar to eq.(\ref{two})
\bea
\mathbb{A}-\mathbb{B}&=&\langle  \kk(1,m)q\rangle [\kk(m+1,n)\kk(1,m)]\langle q\kk(m+1,n)\rangle
\nonumber\\&\times &
\langle  \kk(1,n)q\rangle \langle q^{-}\mid\gamma_{\m}\mid \kk^{-}(1,n)\rangle
\eea
giving
\bea &&
J_{\m}(1+,\cdots,n+)=\frac{1}{2^{\frac{n}{2}}\kappa^{2}(1,n)}\langle  \kk(1,n)q\rangle \langle q^{-}\mid\gamma_{\m}\mid
\kk^{-}(1,n)\rangle P^{n-1}_{1}
\eea
where
\bea
P^{n-1}_{1}=\sum^{n-1}_{m=1}\frac{\langle  \kk(1,m)q\rangle [\kk(m+1,n)\kk(1,m)]\langle q\kk(m+1,n)\rangle }
{\langle\langle q1,mq\rangle\rangle \langle\langle q(m+1),nq\rangle\rangle}
\eea
for which one can show
\bea
P_{1}^{n-1}=\frac{\kappa^{2}(1,n)}{\langle\langle q1,nq\rangle\rangle}
\label{6}
\eea
\noindent
To evaluate $J(1-,2+,\cdots ,n+)$, we have
\bea
J_{\m}(1-)=\e^{-}_{\m}(1,k_{2})=-\frac{\langle 2^{+}\mid\gamma_{\m}\mid 1^{+}\rangle }
{\sqrt{2}[21]}
\eea
\noindent and we can show that $J_{\m}(1-,2+)=0$, and
\bea
J_{\m}(1-2+3+)&=&\frac{1}{\kappa^{2}(1,3)}(-\kappa_{\n}(2,3)\frac{\langle
2^{+}\mid\gamma^{\n} \mid 1^{+}\rangle } {\sqrt{2}[21]}
J_{\m}(2+3+))\nonumber\\&=&\frac{-1}{\kappa^{2}(1,3)}\langle
\kk^{\pm}(2,3)\mm\gamma_{\n}\mm\kk^{\pm}(2,3) \rangle
\frac{\langle 2^{+}\mid\gamma^{\n}\mid 1^{+}\rangle }
{\sqrt{2}[21]} J_{\m}(2+3+)\nonumber\\&&
\eea
Noting (\ref{555}) we get to
\bea
J_{\m}(1-,2+,3+)&=&\frac{-1}{\kappa^{2}(1,3)}
\frac{[\kk(2,3)\ 2]\langle 12\rangle \langle 1\kk(2,3)\rangle
}{\sqrt{2}[21]\langle 12\rangle }
J_{\m}(2+3+)\nonumber\\&=&\frac{-1}{\sqrt{2}} \frac{ \langle
1^{-}\mm(2\sl+3\sl)\mm 2^{-}\rangle\langle 12\rangle
}{\kappa^{2}(1,3)\kk^{2}(1,2)}
J_{\m}(2+3+)\nonumber\\&=&\frac{-1}{\sqrt{2}} \frac{ \langle  1\ 3
\rangle [3\ 2]\rangle\langle 12\rangle }{\kappa^{2}(1,3)\kk^{2}(1,2)} J_{\m}(2+3+)
\eea
The induction conjecture for $l\geq3$ is given by
\bea
J_{\m}(1-2+\cdots l+)=-2^{-\frac{1}{2}}J_{\m}(2+\cdots l+)\sum^{l}_{m=3}\frac{\langle
1m\rangle [m\kk(2,m)]\langle  1\kk(2,m)\rangle }{\kappa^{2}(1,m-1)\kappa^{2}(1,m)}
\eea
this is reduced to
\bea
J_{\m}(1-2+\cdots n+)=-2^{-\frac{1}{2}}J_{\m}(2+\cdots n+)\frac{\langle
1n\rangle [n\kk(1,n)]\langle  1\kk(1,n)\rangle }{\kappa^{2}(1,n-1)\kappa^{2}(1,n)}
\label{1111}
\eea
This is proven in \cite{giele}. Complex conjugation of the currents eq.(\ref{555}) and eq.(\ref{1111}) gives
$J(1-2-\cdots n-)$ and $J(1+2-\cdots n-)$.


\end{document}